\documentclass[twocolumn,amsmath,amssymb,prl,superscriptaddress]{revtex4-1}

\setcitestyle{super}

\usepackage[final,letterspace=-40]{microtype}
\usepackage[colorlinks=true, linkcolor=blue, urlcolor=blue,  citecolor=blue, anchorcolor=blue]{hyperref}

\usepackage{mathpazo} % add possibly `sc` and `osf` options

\usepackage{graphicx}% Include figure files
\usepackage{dcolumn}% Align table columns on decimal point
\usepackage{bm}% bold math
\usepackage{helvet}

\usepackage{caption}
\DeclareCaptionLabelFormat{plain}{Figure #2 $\boldsymbol |$ }

\begin{document}

\title{Lattice topology dictates photon statistics }

\author{H. Esat Kondakci}
\email{esat@creol.ucf.edu}
\author{Ayman F. Abouraddy}
\author{Bahaa E. A. Saleh}
\affiliation{CREOL, The College of Optics \& Photonics, University of Central Florida, Orlando, Florida 32816, USA}

\begin{abstract} \noindent 	Propagation of coherent light through a disordered network is accompanied by randomization and possible conversion into thermal light. Here, we show that network topology plays a decisive role in determining the statistics of the emerging field if the underlying lattice satisfies chiral symmetry. By examining one-dimensional arrays of randomly coupled waveguides arranged on linear and ring topologies, we are led to a remarkable prediction: the field circularity and the photon statistics in  ring lattices are dictated by its parity -- whether the number of sites is even or odd, while the same quantities are insensitive to the parity of a linear lattice. Adding or subtracting a single lattice site can switch the photon statistics from super-thermal to sub-thermal, or vice versa. This behavior is understood by examining the real and imaginary fields on a chiral-symmetric lattice, which form two strands that interleave along the lattice sites. These strands can be fully braided around an even-sited ring lattice thereby producing super-thermal photon statistics, while an odd-sited lattice is incommensurate with such an arrangement and the statistics become sub-thermal. Such effects are suppressed in the limit of high lattice disorder, whereupon Anderson localization arrests the spread of light around the lattice and eliminates topology-dependent phenomena.
\end{abstract}

\small
\maketitle

%\section{Introduction}

\noindent 
Topology, the study of those properties of geometric objects that remain invariant under continuous transformations such as bending or stretching (homeomorphisms \cite{Nakahara1990} in general), has recently entered optics in several guises. First, the development of topological insulators in condensed matter physics  \cite{Qi2011} has inspired the exploration of analogous concepts in the topology of photonic bands in carefully constructed optical structures  \cite{Wang2009, Yablonovitch2009, Rechtsman2013, Rechtsman2013b, Lumer2013a, Zeuner2014}, which offer intriguing possibilities such as one-way propagation and self-healing edge states  \cite{Lu2014}. In an altogether different vein, topological features of the three-dimensional distribution of the optical field in physical space have been investigated, such as the knottedness of scalar wavefronts  \cite{Leach2004, Irvine2008, Irvine2010, Kedia2013} and the emergence of non-trivial topological structure in tightly focused vector fields  \cite{Bauer2015}. A lesser-studied impact of topology on optics, however, is that resulting from the interaction of light with a photonic structure that itself features non-trivial topology. An early prescient study examined optical scattering off knotted configurations to discern the underlying topology  \cite{Manuar1995}. 

\begin{figure*}[!t]
	\includegraphics[scale=1.2]{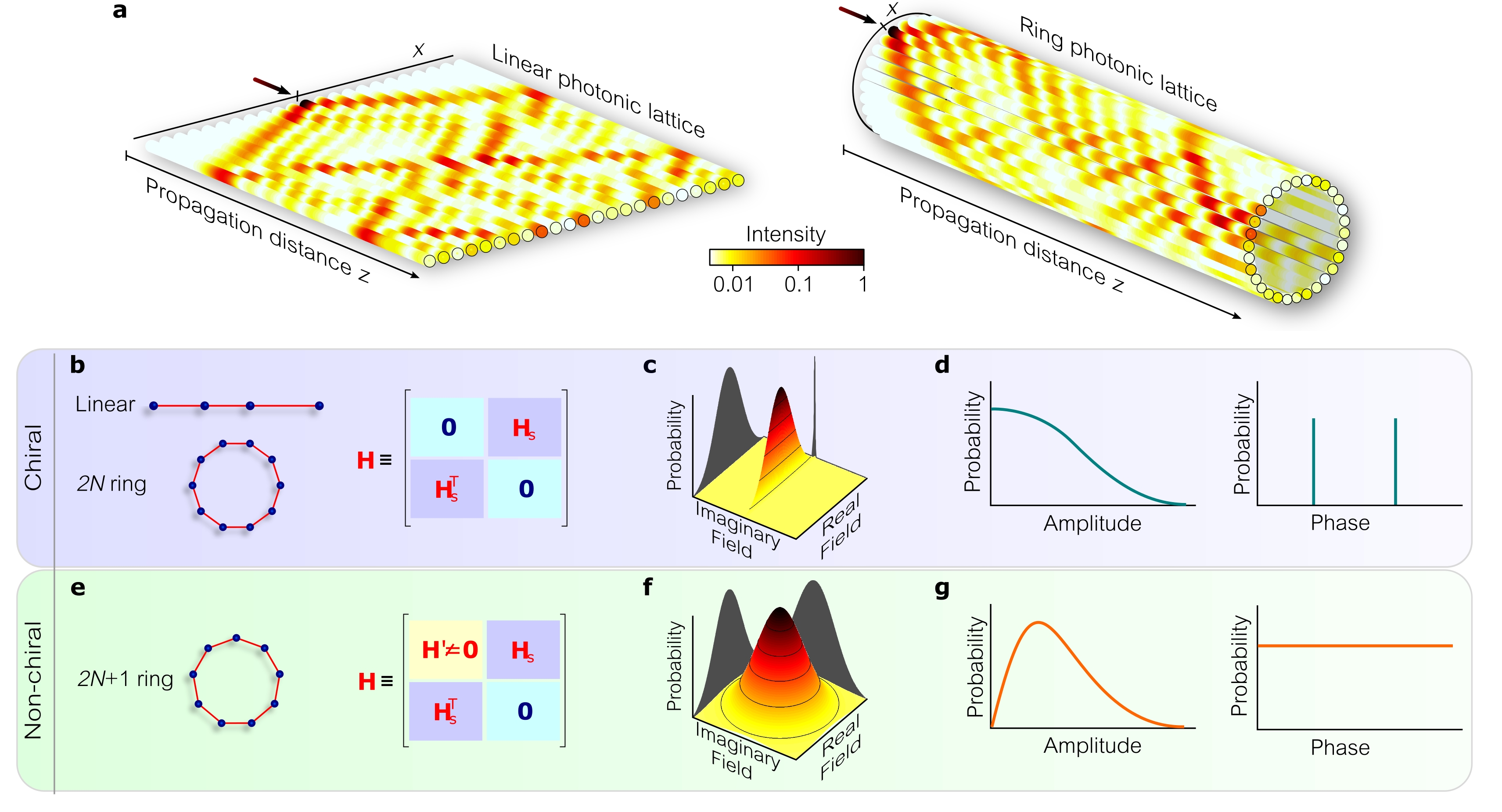}
	\caption{\label{fig:concept} \textbf{Propagation of light along disordered coupled waveguides arranged on different 1D lattice topologies.} \\  \textbf{a}, Evolution of the intensity along single realizations of 1D disordered linear and ring photonic lattices. The black arrow indicates the single waveguide excited at the input. \textbf{b}, The coupling matrix associated with linear or even-sited ring lattices with off-diagonal disorder can be rearranged into block off-diagonal form. \textbf{c}, Probability density  function of the complex field produced by the lattices in (\textbf{b}). The field has only one quadrature component, either real or imaginary. \textbf{d} The probability distribution of the amplitude is half bell-shaped and the phase distribution is discrete. 
	\textbf{e-g}, Similar to (\textbf{b-d}) for odd-sited ring lattices. \textbf{e}, The coupling matrix cannot be rearranged into block off-diagonal form, (\textbf{f}) the probability distribution of the field is circularly symmetric in the complex plane resulting in (\textbf{g}) a Rayleigh-like amplitude distribution and a uniform phase distribution.}
\end{figure*}

Here, we investigate the distinguishing features of bound optical fields propagating along disordered one-dimensional (1D) lattices having distinct underlying topologies -- the line and the ring (Fig.~\ref{fig:concept}(a)). At first sight, it appears that topology should have no impact on the field confined to such a structure and -- moreover -- that any non-trivial topological signatures displayed by the field are likely to be obscured as a result of disorder. Surprisingly, we find that the lattice topology plays a decisive role in determining both the circularity of the field quadratures \textit{and} the photon statistics when a particular disorder-immune lattice symmetry is satisfied -- so-called `chiral symmetry'  \cite{Dyson1953, Gade1991, Gade1993, Bocquet2003a}. The physical platform we examine is an array of parallel waveguides with nearest-neighbor-only evanescent coupling  \cite{Christodoulides2003a}, and we investigate the optical statistics when coherent light excites a single site  \cite{Raedt1989, Schwartz2007a, Lahini2008a, Martin2011a, Segev2013, Kondakci2015, Kondakci2015b}; but the results can be readily extended to other photonic realizations.

Specifically, we show that the photon statistics produced by a disordered ring lattice with chiral symmetry is sensitive to its parity, while linear lattices featuring the same disorder lack this striking characteristic. The traditional concept of periodic boundary conditions -- introduced by Max Born  \cite{Born1912} -- enforces the notion of self-consistency around a ring. In the case of the disordered ring lattices examined here, the delineation of the field into real and imaginary quadratures occupying alternating sites as a result of chiral symmetry brings about a new self-consistency condition: the complete braiding of the two strands representing the field quadratures. Successful braiding is incommensurate with an odd-sited ring lattice and can only be realized on even-sited ring lattices. The photon statistics associated with satisfying this boundary condition are markedly different from those produced in lattices where it is not. Indeed, super-thermal photon statistics are produced by an even-sited ring lattice, and removing or adding a single site from the lattice results in an abrupt change to sub-thermal photon statistics. In linear lattices, these distinctions are entirely absent and the free boundary conditions at the edges of this topology nullify any impact of the lattice parity.

The characteristics of light emerging from such 1D disordered lattices differ from thermal light. The ubiquity of thermal statistics in optics is a consequence of the assumptions underlying the Central Limit Theorem  \cite{Papoulis1965} being readily satisfied under generally common conditions; for example, coherent light scattered from a random surface, or diffused through a disordered medium, acquires thermal statistics upon ensemble averaging  \cite{Goodman2000}. Nevertheless, there are situations for which light exhibits non-Gaussian and/or non-circular statistics  \cite{Goodman1975, Ohtsubo1976, Apostol2005, Jakeman2014}: the quadratures can be Gaussian but not circularly symmetric resulting in Gaussian but non-thermal light  \cite{Picinbono1969, Picinbono1970}; one of the quadratures may be altogether extinguished, resulting in a field having random amplitude but deterministic phase  \cite{Goodman1975, Ohtsubo1976}; or the complex field may be circularly symmetric with two identical and independent but non-Gaussian quadratures  \cite{Apostol2005, Jakeman2014}. We show that all these scenarios are spanned by light emerging from disordered 1D lattices in different topologies that satisfy chiral symmetry.

\section{Results}
\subsection{Chiral-symmetric lattice model} 
We consider coupled identical waveguides with nearest-neighbor-only interactions arranged on two different lattice topologies: the \textit{linear} lattice and the closed \textit{ring} lattice (Fig.~\ref{fig:concept}(a)), the former of which has been studied extensively  \cite{Christodoulides2003a, Lahini2008a, Lahini2009, Thompson2010a, Peruzzo2010b, Lahini2010a, Lahini2011a, Kottos2011, Ghosh2011, Martin2011a, Abouraddy2012b, DiGiuseppe2013, Segev2013, Kondakci2015, Kondakci2015b,Kondakci2016a}. The complex envelope $\mathbf{A}=\{A_x(z)\}_x$ of a coherent monochromatic field at lattice site $x$ evolves according to the matrix equation  \cite{Christodoulides2003a} $i\mathrm{d}\mathbf{A}/\mathrm{d}z+\mathbf{\hat{H}}\mathbf{A}\!=\!0$, where $\mathbf{\hat{H}}$ is the coupling matrix or the system's Hamiltonian (Fig.~\ref{fig:concept}(b),(e); Appendix) and $z$ is the axial position. We introduce disorder into the lattice by randomizing the waveguide couplings, which may be achieved by varying their separation -- so-called off-diagonal disorder \cite{Soukoulis1981a,Szameit2010,Martin2011a}. We assume random coupling coefficients described by a uniform probability distribution with mean $\bar{C}$ and half-width $\Delta C$. The coupling coefficients are reported in units of $\bar{C}$ such that $0\!\leq\!\Delta C\!\leq\!1$, and the axial position $z$ is in units of the coupling length $\ell=1/\bar{C}$.

The field is best described in terms of the eigenvectors $\{\phi_{n}(x)\}_{x}$ and corresponding eigenvalues $b_n$ of $\mathbf{\hat{H}}$, $\mathbf{\hat{H}}\varphi_{n}(x)\!=\!b_{n}\varphi_{n}(x)$, which are all real-valued since $\mathbf{\hat{H}}$ is real and symmetric. If the array is excited at $z\!=\!0$ by a field $\mathbf{A}(0)\!=\!\{A_x(0)\}\!=\!\sum_{n}\!c_{n}\varphi_{n}(x)$, where $c_{n}$ is the excitation amplitude of the $n^{\mathrm{th}}$ mode, then $A_x(z)=\sum_{n} c_{n} \phi_n(x) e^{ib_n z}$. Such a phasor sum leads one to expect that light emerges at the output with \textit{thermal} statistics and \textit{circular symmetry} since $\{b_{n}\}$, $\{\varphi_{n}(x)\}$, and $\{c_{n}\}$ are all random variables in the probability space of the statistical ensemble. Thermal light is characterized by a complex optical field whose real and imaginary quadrature components are statistically independent and identically distributed Gaussian random variables. The field statistics thus exhibit circular symmetry in the complex plane  \cite{Goodman2000, Saleh1978} and the field amplitude is Rayleigh-distributed with a uniform phase distribution, resulting in an intensity $I$ that is exponentially distributed with normalized variance $\textrm{Var}(I)\big/\!\langle I\rangle^2\!=\!1$; Fig.~\ref{fig:concept}(f),(g). We proceed to show that, surprisingly, thermal statistics are \textit{not} necessarily produced upon traversing a disordered lattice of coupled waveguides.

\begin{figure*}[!t]
	\includegraphics[scale=1.2]{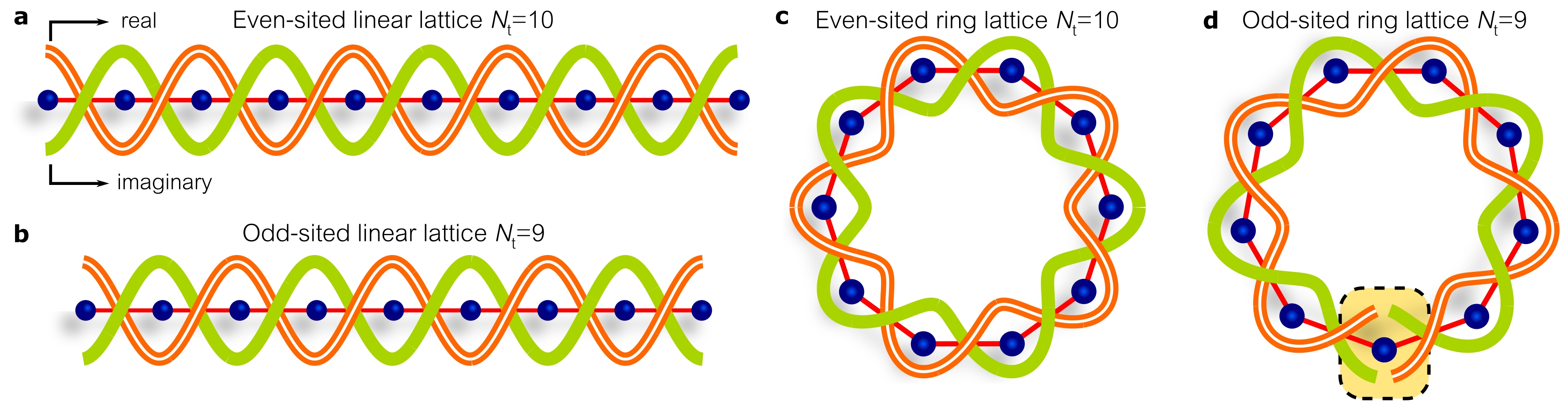}
	\caption{\label{fig:braiding} \textbf{Braiding of the real and imaginary field components around linear and ring lattices endowed with chiral symmetry.} \\ \textbf{a},\textbf{b}, Braiding the real and imaginary field components along linear lattices is insensitive to the lattice parity -- whether (\textbf{a}) even-sited or (\textbf{b}) odd-sited -- due to the free lattice boundaries. \textbf{c}, Braiding is complete around an even-sited ring lattice. \textbf{d}, Braiding cannot be completed around an odd-sited ring lattice. Incommensurability breaks the chiral symmetry.}
\end{figure*}

\begin{figure*}[!t]
	\includegraphics[scale=1.2]{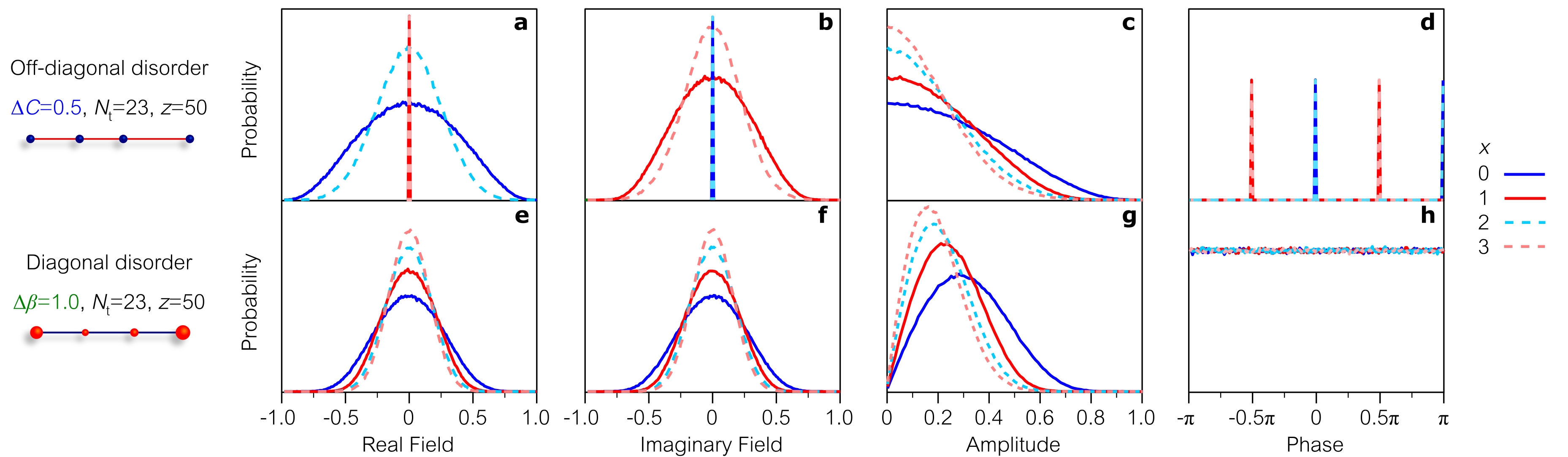}
	\caption{\label{fig:FieldLinear} \textbf{Probability distributions of the field quadratures, amplitude, and phase in linear disordered lattices.} \\ \textbf{a-d,} Probability distributions of the (\textbf{a}) real quadrature, (\textbf{b}) imaginary quadrature, (\textbf{c}) amplitude, and (\textbf{d}) phase of the complex optical field at different output sites $x$ after propagating through a photonic lattice with off-diagonal disorder. The real (imaginary) quadrature vanishes on the odd (even) sites, and the phase takes binary values with equal probability: $0$ or $\pi$ ($\pm\pi/2$) at even (odd) positions. \textbf{e-h,} Same as in (\textbf{a-d}) for a lattice with diagonal disorder. The field is circularly symmetric, resulting in a uniform phase distribution. We use throughout ensembles of $10^5$ realizations and a single lattice site ($x\!=\!0$) is excited.}
\end{figure*}

Linear lattices with off-diagonal disorder display a disorder-immune symmetry since $\mathbf{\hat{H}}$ can be rearranged in a block off-diagonal form (Fig.~\ref{fig:concept}(b)) -- so-called `chiral symmetry'  \cite{Bocquet2003a}. In this case, the eigenvalues and eigenvectors of $\mathbf{\hat{H}}$ come in skew-symmetric pairs, $b_{-n}\!=\!-b_{n}$ and $\phi_{-n}(x)\!=\!(-1)^x\phi_{n}(x)$, in \textit{every} realization -- not only on average -- and for \textit{every} lattice disorder level $\Delta C$, such that the output field for unit input excitation at $x\!=\!0$ is
\begin{equation}\label{eq:fieldoffdiag}
A_x(z)=\begin{cases}2 \sum\limits_{n} \phi_{n}(0) \phi_{n}(x)  \cos(b_n z),& x\mbox{ even,}\\ 2i\sum\limits_{n} \phi_{n}(0) \phi_{n}(x)  \sin(b_n z), & x\mbox{ odd}. \end{cases}
\end{equation}
The complex envelope $A_x(z)$ thus alternates between \textit{real} and \textit{imaginary} values at \textit{even}- and \textit{odd}-indexed waveguides, respectively. Because only one quadrature survives at any lattice site, the phase takes on only \textit{discrete} values, 0 or $\pi$ ($\pm\pi\!/2$) at even-indexed (odd-indexed) sites; see Appendix. Consequently, the emerging light in general is \textit{non}-thermal. For example, once the steady state is reached in the low-disorder limit, one field quadrature is approximately Gaussian and the intensity has a chi-square probability distribution with one degree of freedom  \cite{Saleh1978} $P(I)=\exp(-I/2\langle I \rangle)\big/(2\pi\langle I \rangle I)^{1/2}$ characterized by a normalized variance $\textrm{Var}(I)\big/\!\langle I \rangle^2\!=\!2$; Fig.~\ref{fig:concept}(c),(d).  Furthermore, associated with the topology-dependent circularity of the field quadratures is a concomitant dependence of the photon statistics on lattice topology. For this purpose, we take the normalized intensity correlation, $g^{(2)}\!=\!\langle I^2\rangle\big/\langle I\rangle^2\!=\!\mathrm{Var}(I)\big/\langle I \rangle^2 +1$, which is a standard descriptor for the field randomness; $\langle\cdot\rangle$ denotes ensemble averaging and $I$ is the intensity.

Any classical field can be described in terms of photon-number statistics. The transformation from intensity statistics to photon-number distribution can be obtained via Mandel's formula given by 
\begin{equation}
\mathrm{P}(n_\mathrm{ph})=\int_0^\infty \frac{\mu^{n_\mathrm{ph}}}{n_\mathrm{ph}!} e^{-\mu} \mathrm{P}(I)\mathrm{d}I, 
\end{equation}
where $n_\mathrm{ph}$ is the photon number and $\mu = \langle n_\mathrm{ph}\rangle$ is the average photon number proportional to the intensity  \cite{Mandel1995}. In this case, the normalized intensity correlation becomes $g^{(2)}=\langle n_\mathrm{ph}(n_\mathrm{ph}-1)\rangle/\langle n_\mathrm{ph}\rangle^2$ and is related to Mandel's $Q$-parameter via $Q=\mathrm{Var}(n_\mathrm{ph})/\langle n_\mathrm{ph}\rangle-1=\langle n_\mathrm{ph} \rangle (g^{(2)}-1)$.

\subsection{From linear to ring topology} 
Equation~(\ref{eq:fieldoffdiag}) provides a hint for the impact of lattice topology on the field circularity and photon statistics. The field at any axial position along a linear waveguide array endowed with chiral symmetry can be viewed as a result of \textit{braiding} two different strands occupying adjacent sites from one end of the lattice to the other: a `real' strand on even-indexed sites and an `imaginary' strand on odd-indexed sites; Fig.~\ref{fig:braiding}(a),(b). The free boundaries of the linear array make this braiding insensitive to the number of sites on the lattice. Folding a linear lattice into a ring gives rise to two scenarios that depend decisively on the ring-\textit{parity} -- whether it has an even or odd number of sites. If the ring lattice is \textit{even}-sited, the real and imaginary strands form two \textit{closed braided} rings (Fig.~\ref{fig:braiding}(c)), which preserves the chiral symmetry. The emerging light thus remains non-circular as in its linear counterpart. On the other hand, if the ring is \textit{odd}-sited, the lattice structure is incommensurate with closed braided real and imaginary strands (Fig.~\ref{fig:braiding}(d)). Here, chiral symmetry is in fact broken, and the field may emerge with \textit{circular} symmetry.

\begin{figure*}[!t]
	\includegraphics[scale=1.2]{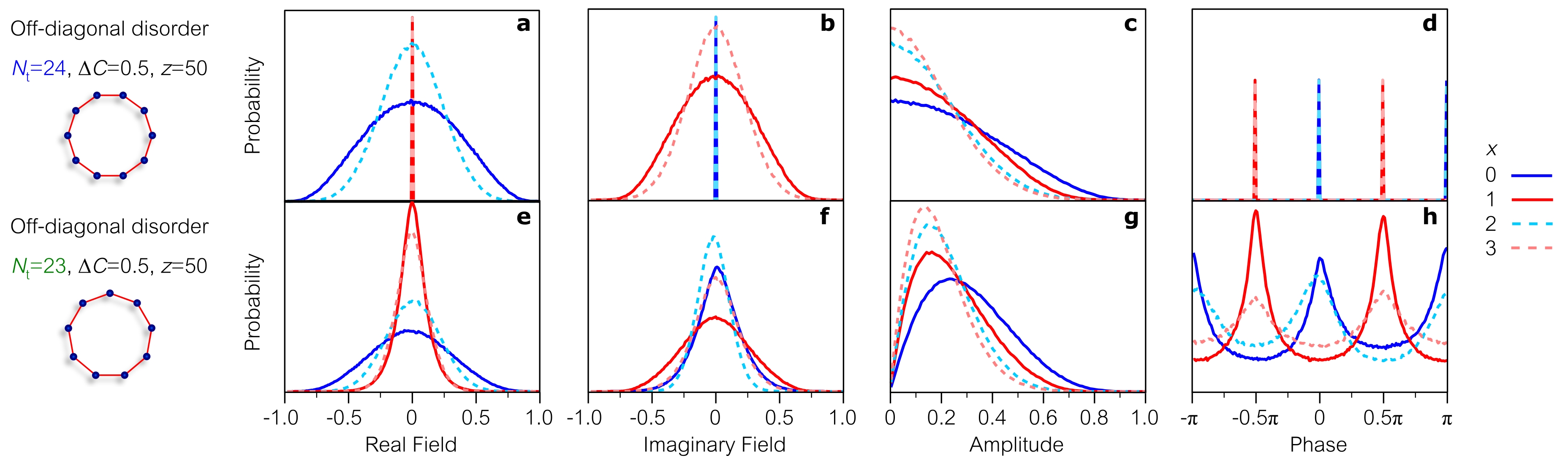}
	\caption{\label{fig:FieldRing} \textbf{Probability distributions of the field quadratures, amplitude, and phase in ring lattices with off-diagonal disorder.} \\ \textbf{a-d,} Probability distributions of the (\textbf{a}) real quadrature, (\textbf{b}) imaginary quadrature, (\textbf{c}) amplitude, and (\textbf{d}) phase of the complex optical field at different output sites $x$ after propagating through an \textit{even}-sited ring lattice. The real (imaginary) quadrature vanishes on the odd (even) sites, and the phase takes binary values with equal probability: $0$ or $\pi$ ($\pm\pi/2$) at even (odd) positions. \textbf{e-h,} Same as in (\textbf{a-d}) for an \textit{odd}-sited ring lattice. \textbf{g,} Amplitudes of the fields take Rayleigh-like distributions. \textbf{h,} Phase distributions reflect an intermediate stage between discrete and uniform probability distributions.}
\end{figure*}

\begin{figure*}[!t]
	\includegraphics[scale=1.2]{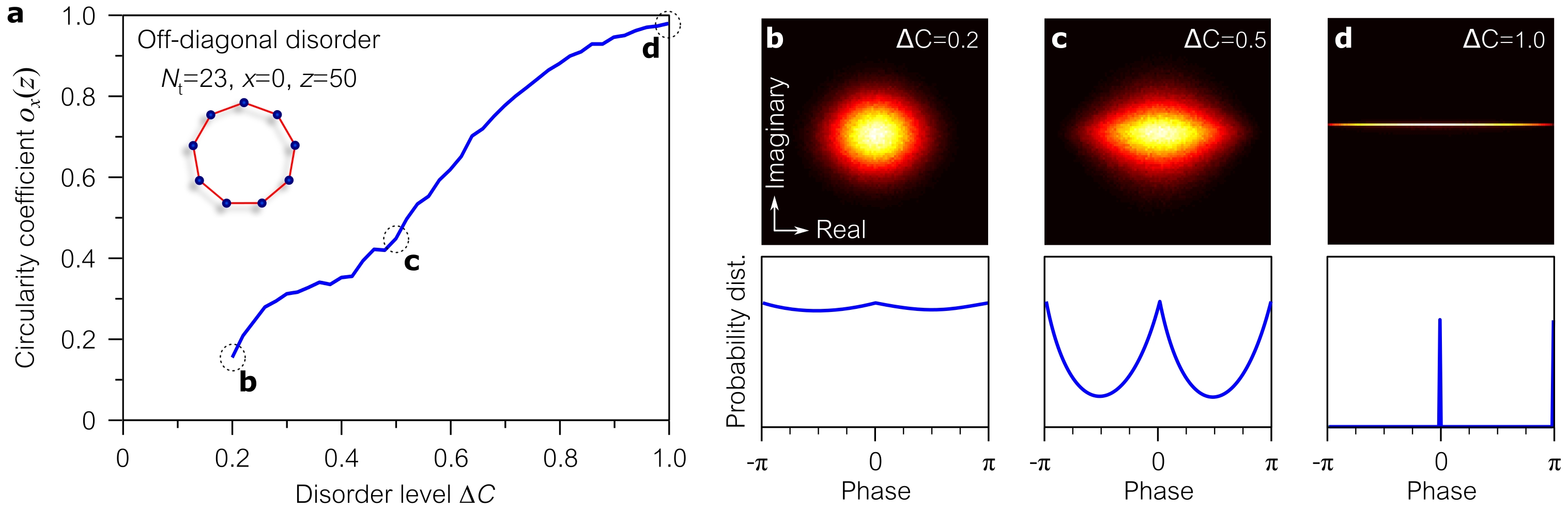}
	\caption{\label{fig:Circularity} \textbf{Circularity in odd-sited ring lattices with off-diagonal disorder.} \\ \textbf{a}, Increasing the disorder level $\Delta C$ reduces the circularity of the output field at the excitation waveguide ($x\!=\!0$) and increases the circularity coefficient $o_{x}(z)$. \textbf{b}-\textbf{d}, The field probability distribution in the complex plane for selected disorder levels. Bright colors represent higher probability density. The second row gives the probability distribution of the corresponding field phase.}
\end{figure*}

We are thus led to a remarkable prediction: the circularity of the field quadratures and the photon statistics of light emerging from a disordered ring-lattice \textit{is sensitive to the ring's parity}. In the steady state, this can entail a dramatic change in the normalized variance by simply adding or removing \textit{a single} waveguide from the ring, in contradistinction to the \textit{in}sensitivity of the optical statistics in a disordered linear lattice to its parity. 

Simulations of linear and ring lattices when excited from a single site confirm these predictions. An example of the statistics produced by a disordered chiral-symmetric linear lattice is provided in Fig.~\ref{fig:FieldLinear}(a)-(d). One field quadrature has a bell-shaped probability distribution while the other is deterministic. Consequently, the probability distribution for the phase $\phi$ is discrete and the field is not circularly symmetric. In the case of a disordered linear lattice lacking chiral symmetry, e.g. so called `diagonal disorder' (Appendix), the field quadratures have identical distributions and the phase $\phi$ is uniformly distributed over the range [$0,2\pi$], so that the field phasor is indeed circularly symmetric (Fig.~\ref{fig:FieldLinear}(e)-(h)).

In ring lattices with off-diagonal disorder, the field statistics depend crucially on the ring parity. When the ring is even-sited (here, the number of sites is $N_\mathrm{t}\!=\!24$), the field quadratures are braided around the ring (Fig.~\ref{fig:braiding}(c)) and the simulations confirm our prediction that the discrete phase distribution -- observed in the corresponding linear lattice -- is maintained (compare Fig.~\ref{fig:FieldRing}(a)-(d) to Fig.~\ref{fig:FieldLinear}(a)-(d)). However, when the ring is odd-sited ($N_\mathrm{t}\!=\!23$), the distribution of $\phi$ is no longer discrete (compare Fig.~\ref{fig:FieldRing}(e)-(h) to Fig.~\ref{fig:FieldLinear}(e)-(h)), but instead shows an intermediate stage between the discrete and uniform distributions observed in disordered linear lattices endowed with and lacking chiral symmetry, respectively, therefore signifying a departure from field circularity. Indeed, simulations show that $\phi$ has a discrete distribution when $\Delta C\!\rightarrow\!1$ (the fields lacks circular symmetry) and a uniform distribution when $\Delta C\!\rightarrow\!0$ (the field is circularly symmetric).

\begin{figure*}[t!]
	\includegraphics[scale=1.08]{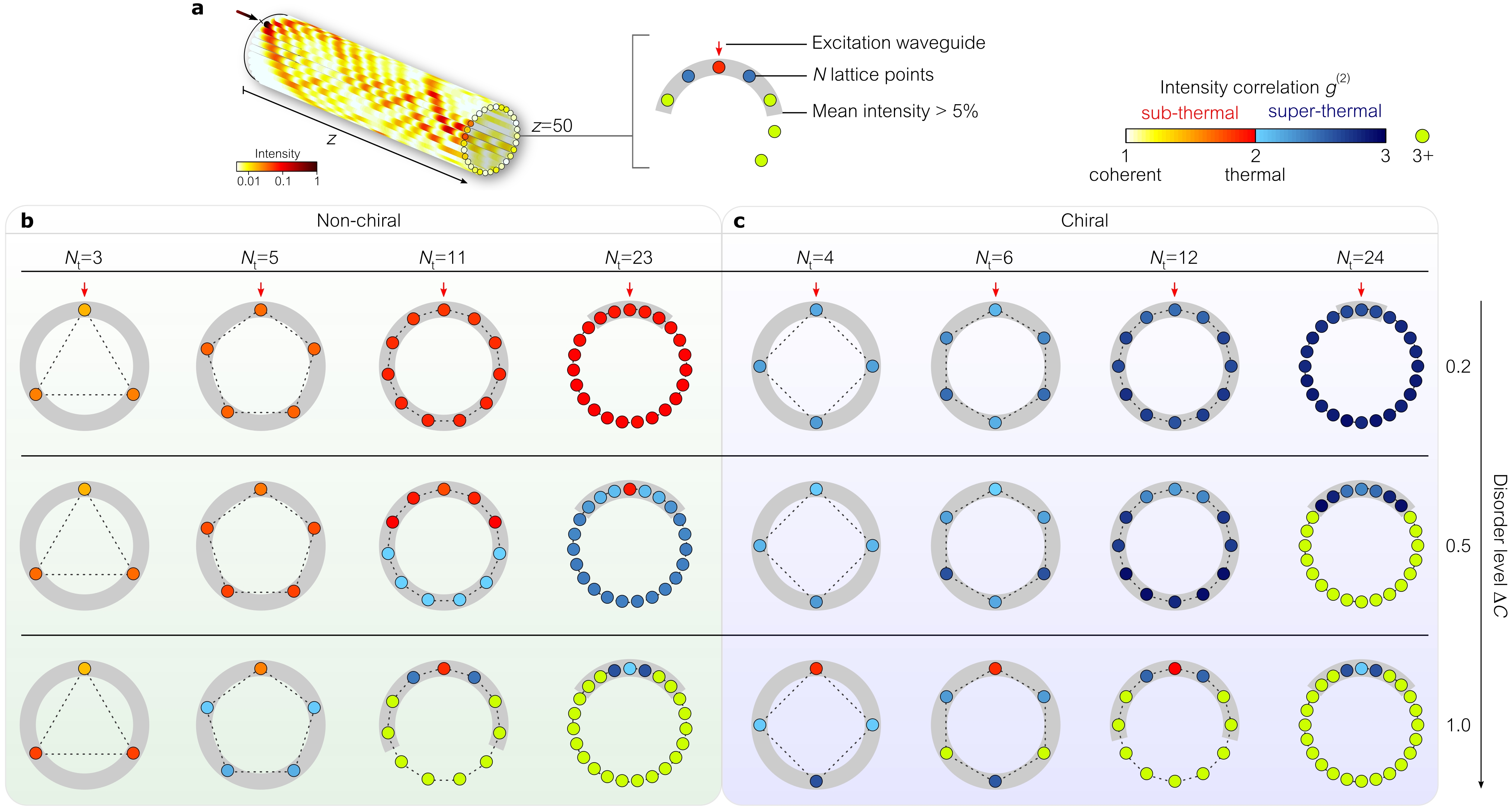}
	\caption{\label{fig:g2} \textbf{Normalized intensity correlation $\mathbf{g^{(2)}}$ in ring lattices with off-diagonal disorder.} \\ \textbf{a}, Evolution of the intensity on a ring lattice in a single realization of disorder. The arrow identifies the input site. \textbf{b}, $g^{(2)}$ at the output of odd-sited ring lattices, which  lack chiral symmetry. The color in the small  circles (corresponding to the lattice points) depicts the normalized intensity correlation $g^{(2)}$. The field statistics are mainly sub-thermal. In the off-set waveguides the statistics may become super-thermal, but this comes with a reduction in intensity. The two regimes of sub-thermal and super-thermal photon statistics are denoted by different color palettes. The red arrow identifies the excitation site. The gray shading encompasses the lattice sites with the mean intensity exceeds $5\%$ of the input power. \textbf{c}, $g^{(2)}$ in even-sited ring lattices with chiral symmetry. The statistics are mainly super-thermal. At offset waveguides, $g^{(2)}$ may take very high values at the expense of low intensity.}
\end{figure*}

The dependence of the field circularity on the disorder level $\Delta C$ can be placed on a quantitative basis using the concept of the circularity quotient $\mathrm{circ}(w_1)$ of a complex random variable $w_1$, which is defined as the ratio between the variable's pseudo-variance and variance  \cite{Picinbono1994,Ollila2008}, $\mathrm{circ}\{w_1\}\!=\!\mathrm{cov}(w_1,w_1^*)\big/\mathrm{cov}(w_1,w_1)$,
where $\mathrm{cov}(w_1,w_2)\!=\!\langle w_1 w_2^*\rangle$. It is clear that $\mathrm{circ}\{w_1\}\!=\!0$ whenever $w_1$ is a circularly symmetric complex random variable, while $|\mathrm{circ}\{w_1\}|\!=\!1$ for a linear $w_1$, so we quantify the field circularity by $o_x(z)\!=\!|\mathrm{cir}\{A_x(z)-\langle A_x(z)\rangle\}|$. In any even-sited chiral ring lattice, the circularity coefficient $o_x(z)\!=\!1$ for all disorder levels. In an odd-sited ring lattice, $o_{x}(z)$ increases with $\Delta C$ as shown in Fig.~\ref{fig:Circularity}.

\subsection{Lattice topology and Anderson localization}
The dependence of the field circularity on $\Delta C$ in an odd-sited ring is a consequence of the transverse Anderson localization of light along the waveguides  \cite{Anderson1958a}. At small disorder levels ($\Delta C\!\rightarrow\!0$), the field spreads around the whole ring. The odd number of sites is incommensurate with fully braiding the real and imaginary strands of the field, thereby breaking the chiral symmetry and leading to a circularly symmetric field and $o_{x}(z)\!\rightarrow\!0$. On the other hand, light is localized in the vicinity of the excitation site $x\!=\!0$ for large disorder levels ($\Delta C\!\rightarrow\!1$) and hence does not extend around the ring lattice. As such, the lattice effectively corresponds to a linear lattice in which the braiding of the real and imaginary field strands is always possible, chiral symmetry is preserved, the field is no longer circular, and ultimately $o_{x}(z)\!\rightarrow\!1$. In an even-sited ring lattice, whether light is localized or not, the field strands can always be braided and $o_{x}(z)\!=\!1$ for all disorder values $\Delta C$.

In addition, Anderson localization prevents the realization of Gaussian statistics, especially for high disorder levels. When light is strongly localized, it is coupled to a small number of lattice modes whose excitation amplitudes thus become strongly correlated, which implies the non-validity of the central limit theorem. Consequently, the field statistics deviate from the Gaussian distribution, although they remain bell-shaped. Reducing $\Delta C$ reverses this trend: the excitation is coupled to a larger number of lattice modes with independent amplitudes and Gaussian statistics are produced.

\subsection{Interplay of ring parity, disorder level, and lattice size} 
The interplay between disorder level and lattice size in determining $g^{(2)}$ in even- and odd-sited ring lattices is illustrated in Fig.~\ref{fig:g2}. Focusing first on $g^{(2)}$ at the excitation site $x\!=\!0$, a marked distinction appears immediately between odd- and even-sited ring lattices: in the former $1\!<\!g^{(2)}\!\lesssim\!2$, corresponding to sub-thermal photon statistics; while $2\!\lesssim\!g^{(2)}\!<\!3$ in the latter, corresponding to super-thermal statistics. Consider the first row in Fig.~\ref{fig:g2}(b),(c) corresponding to a low disorder level ($\Delta C\!=\!0.2$). In comparing lattices with 23 and 24 sites, we observe that the addition of a \textit{single} site to the ring lattice results in a jump in $g^{(2)}$ from 2 to 3. In a ring lattice with 23 sites, the real and imaginary field strands cannot be consistently braided. Consequently, the two quadratures are symmetric and the low $\Delta C$ results in approximately Gaussian statistics. The field is circularly symmetric $o_{0}(z)\!\approx\!0.1$ and thermal $g^{(2)}\!\approx\!2$ (Fig.~\ref{fig:g2}(b)). With the addition of a single site, the real and imaginary strands \textit{can} be consistently braided around the 24-sited ring lattice. The field is no longer circularly symmetric $o_{0}(z)\!=\!1$, and one quadrature has approximately Gaussian statistics leading to super-thermal light with $g^{(2)}\!\approx\!3$ (Fig.~\ref{fig:g2}(c)). In addition to the parity-based demarcation of the light statistics, the lattice size also effects the limit of thermalization. For a fixed disorder level, reducing the lattice size \textit{lowers} $g^{(2)}$ in both even- and odd-sited lattices.

When the disorder level is low, $g^{(2)}$ takes similar values at all lattice sites since the intensity is evenly distributed. As the disorder level increases, transverse localization of light dominates and statistics become non-uniform across the lattice. Increasing the disorder level has the effect of blurring the distinction between the even- and odd-sited lattices. This may be understood by observing that localization associated with increasing disorder diminishes the extent to which light spreads across the lattice. In the limit of high disorder, light remains confined around the excitation site, thereby nullifying the impact of lattice size and thus its parity. Therefore, there is little contrast between the photon statistics produced in lattices with different parities when the lattice size is larger than the transverse localization width. For all lattice sizes in this case, $g^{(2)}\!\approx\!2$ on the excitation site. Finally, $g^{(2)}$ generally increases in off-center lattice sites $x\!\neq\!0$, especially in the limit of high disorder where localization is most pronounced. This is associated, nevertheless, with extremely low intensity levels (denoted by green lattice sites in Fig.~\ref{fig:g2}(b),(c) where $g^{(2)}\!>\!3$).

\section{Discussion}

We have found that topology plays an unexpected role in determining the thermalization statistics of light propagating in a disordered lattice of coupled waveguides. In chiral linear lattices, $g^{(2)}$ always corresponds to super-thermal statistics ($g^{(2)}>2$), while $g^{(2)}$ in a ring lattice depends on the ring \textit{parity} -- whether it is even- or odd-sited. Adding or removing a single lattice site can produce a dramatic shift in $g^{(2)}$ from super-thermal to sub-thermal statistics.

Although the construction of optical structures with non-trivial topology is challenging, recent advances in the precise fabrication of coupled waveguide arrays  \cite{Meany2015}, modulated fiber loops \cite{Regensburger2011, Regensburger2012}, liquid crystals  \cite{Martinez2014, Copar2015a}, and on-chip coupled resonators  \cite{Mookherjea2008, Mookherjea2014} provide routes for producing optical structures in which the effects predicted here may be observed.

In this study, we have focused on the scenario where a single lattice is coherently excited. Previous studies of disordered linear lattices indicate that the illumination configuration can modify the photon statistics in crucial ways. Indeed, it has been shown theoretically  \cite{Kondakci2015b} and experimentally  \cite{Kondakci2016} that changing the relative phase between two excited sites can help tune the photon statistics between the two extremes of sub-thermal and super-thermal statistics. We expect that the exploration of illumination configurations on different lattice topologies may yet reveal further surprises.

\appendix*
\section{Appendix}

\newcommand{\db}{\Delta\beta_} 

For both linear and ring lattices, the propagation of an optical field is described by the first-order differential equation  \cite{Raedt1989, Christodoulides2003a} 
\begin{equation*}
	i\frac{\mathrm{d}E_{x}}{\mathrm{d}z}+\beta_{x} E_{x}+C_{x,x-1} E_{x-1}+C_{x,x+1} E_{x+1}=0,
\end{equation*}
where $E_{x} (z)$ is the complex optical field in the ${x}^{\mathrm{th}}$ waveguide at the axial position $z$, $\beta_x$ is the propagation constant in the $x^{\mathrm{th}}$ waveguide, and $C_{x,x+1}=C_{x+1,x}$ is the coupling coefficient between waveguides $x$ and $x+1$. If $\bar{\beta}$ is the mean propagation constant, $\beta_x=\bar{\beta}+\Delta\beta_x$, and $A_x(z)$ is the complex field envelope, then substituting $E_x(z)=A_x(z)\exp(i\bar{\beta}z)$ into the evolution equation produces the matrix equation 
\begin{equation*}
	i\frac{\mathrm{d}\mathbf{A}}{\mathrm{d}z}+\mathbf{\hat{H}}\mathbf{A}=0,
\end{equation*}
where $\mathbf{A}=\{A_x(z)\}_x$ and $\mathbf{\hat{H}}$ is the coupling matrix or the system's Hamiltonian.

A signature distinguishing matrices with chiral ensembles is that they can be transformed into block off-diagonal form \cite{Gade1991,Gade1993}. The necessary transformation is performed by grouping the even- and odd-indexed sites. Examples of the matrices and their transformed versions for various cases are given here:

\vspace*{0.2cm}
\textit{Example 1} -- Linear lattices with \textit{diagonal} disorder and $N_\mathrm{t}\!=\!5$. Here $\mathbf{\hat{H}}$ cannot be transformed into the block-diagonal form, and the lattice thus lacks chiral symmetry.
\begin{equation*}\label{eq:matrix2}
	\begin{split}
		\mathbf{\hat{H}}=
		\begin{pmatrix}
			\db{0}	&\bar{C}&0		&0		&0 		\\
			\bar{C}	&\db{1}	&\bar{C}&0		&0 		\\
			0		&\bar{C}&\db{2}	&\bar{C}	&0 		\\
			0		&0		&\bar{C}&\db{3}	&\bar{C}\\
			0		&0		&0		&\bar{C}&\db{4}
		\end{pmatrix} \\
		\rightarrow
		\begin{pmatrix}
			\db{0}	&0		&0		&\bar{C}&0 		\\
			0		&\db{2}	&0		&\bar{C}&\bar{C}\\
			0		&0		&\db{4}	&0		&\bar{C}\\
			\bar{C}&\bar{C}&0		&\db{1}	&0		\\
			0		&\bar{C}&\bar{C}&0		&\db{3}
		\end{pmatrix}
	\end{split}
\end{equation*}

\vspace*{0.2cm}
\textit{Example 2} -- Linear lattices with \textit{off-diagonal} disorder and $N_\mathrm{t}=5$. Here $\mathbf{\hat{H}}$ \textit{can} be transformed into the block-diagonal form, indicating the presence of chiral symmetry.
\begin{equation*}\label{eq:matrix3}
	\begin{split}
		\mathbf{\hat{H}}=
		\begin{pmatrix}
			0		&C_{0,1}&0		&0		&0 		\\
			C_{1,0}	&0		&C_{1,2}&0		&0 		\\
			0		&C_{2,1}&0		&C_{2,3}&0 		\\
			0		&0		&C_{3,2}&0		&C_{3,4}\\
			0		&0		&0		&C_{4,3}&0
		\end{pmatrix} \\
		\rightarrow
		\begin{pmatrix}
			0		&0		&0		&C_{0,1}&0 		\\
			0		&0		&0		&C_{2,1}&C_{2,3}\\
			0		&0		&0		&0		&C_{4,3}\\
			C_{1,0}	&C_{1,2}&0		&0		&0		\\
			0		&C_{3,2}&C_{3,4}&0		&0
		\end{pmatrix}
	\end{split}
\end{equation*}

In these two cases, $N_\mathrm{t}$ does not play a role in the transformation. In the case of ring lattices with off-diagonal disorder, the outcome surprisingly depends on $N_\mathrm{t}$.

\vspace*{0.2cm}
\textit{Example 3} -- Even-sited ring lattice with off-diagonal disorder ($N_\mathrm{t}=6$). The block-diagonal form signals the presence of chiral symmetry and hence the eigenmodes come in chiral-symmetric pairs.
\begin{equation*}\label{eq:matrix4}
	\begin{split}
		\mathbf{\hat{H}}=
		\begin{pmatrix}
			0		&C_{0,1}&0		&0		&0  & C_{0,5}\\
			C_{1,0}	&0		&C_{1,2}&0		&0 & 0		\\
			0		&C_{2,1}&0		&C_{2,3}&0 	&0	\\
			0		&0		&C_{3,2}&0		&C_{3,4} & 0\\
			0	&0		&0		&C_{4,3}&0 & C_{4,5} \\
			C_{5,0} & 0 & 0 & 0 & C_{5,4} & 0
		\end{pmatrix} \\
		\rightarrow
		\begin{pmatrix}
			0		&0		&0     &C_{0,1}&0 	& C_{0,5}	\\
			0		&0		&0		&C_{2,1}&C_{2,3} & 0\\
			0      &0		&0		&0		&C_{4,3} & C_{4,5}\\
			C_{1,0}	&C_{1,2}&0 &0		&0	& 0	\\
			0		&C_{3,2}&C_{3,4}&0		&0 & 0 \\
			C_{5,0} & 0 & C_{5,4} & 0 & 0 & 0
		\end{pmatrix}
	\end{split}
\end{equation*}

\vspace*{0.2cm}
\textit{Example 4} -- Odd-sited ring lattices with off-diagonal disorder ($N_\mathrm{t}=5$). Although the lattice is characterized by off-diagonal disorder, because the lattice is odd-sited, the block-diagonal form is not obtained and the chiral symmetry is thus lacking. The off-diagonal blocks in the matrix are interacting through matrix elements $C_{4,0}$ and $C_{0,4}$, which are equal.
\begin{equation*}\label{eq:matrix5}
	\begin{split}
		\mathbf{\hat{H}}=
		\begin{pmatrix}
			0		&C_{0,1}&0		&0		&C_{0,4}\\
			C_{1,0}	&0		&C_{1,2}&0		&0 		\\
			0		&C_{2,1}&0		&C_{2,3}&0 		\\
			0		&0		&C_{3,2}&0		&C_{3,4}\\
			C_{4,0}	&0		&0		&C_{4,3}&0
		\end{pmatrix} \\
		\rightarrow
		\begin{pmatrix}
			0		&0		&C_{0,4}&C_{0,1}&0 		\\
			0		&0		&0		&C_{2,1}&C_{2,3}\\
			C_{4,0}	&0		&0		&0		&C_{4,3}\\
			C_{1,0}	&C_{1,2}&C_{1,4}&0		&0		\\
			0		&C_{3,2}&C_{3,4}&0		&0
		\end{pmatrix}
	\end{split}
\end{equation*}

In the case of diagonal disorder  \cite{Anderson1958a, Lahini2008a}, we assume random propagation constants described by a zero-mean uniform probability distribution of half-width $\Delta \beta$ (Fig.~\ref{fig:concept}(b)) $0\!\leq\!\Delta\beta\!\leq\!3$

Since the field is circularly symmetric and takes random independent phases at the sites on a lattice lacking chiral symmetry, folding this lattice into a closed ring topology will have no impact on the field. In linear lattices with \textit{diagonal} disorder, the complex field is indeed circularly symmetric at sufficiently large $z$ such that $\mathrm{Var}(b_{n})z\!\gg\!2\pi$, where $\mathrm{Var}(b_{n})$ is the variance in the eigenvalues $b_n$. Furthermore, $A_{x}(z)$ is thermal with circularly symmetric gaussian quadratures for a wide range of disorder levels $\Delta\beta$.

\bibliography{library_short,books}
\end{document}